\documentclass[twocolumn]{elsart}

\usepackage{natbib}

 \usepackage{epsfig}

\usepackage{amssymb}

\newcommand{\Msun}{~M_\odot}

\newcommand{\kms}{\rm ~km~s^{-1}}

\newcommand{\ml}{~\Msun ~\rm yr^{-1}}

\newcommand\lsim{\raise0.3ex\hbox{$<$}\kern-0.75em{\lower0.65ex\hbox{$\sim$}}}
\newcommand\gsim{\raise0.3ex\hbox{$>$}\kern-0.75em{\lower0.65ex\hbox{$\sim$}}}

\begin{document}
\runauthor{Chevalier}
\begin{frontmatter}



\title{What Gamma-Ray Bursts Explode Into}

\author{Roger A. Chevalier} 
\ead{rac5x@virginia.edu}
\address{Dept. of Astronomy, University of Virginia, P.O. Box 400325,
Charlottesville, VA 22904, USA}

\begin{abstract}
The association of 
long gamma-ray bursts (GRBs) with Type Ib/c supernovae implies
that they explode into the winds
of their Wolf-Rayet progenitor stars.
Although the evolution of some GRB afterglows is consistent with
expansion into a free wind, there is also good evidence for expansion into
a constant density medium.
The evidence includes the evolution of X-ray afterglows (when X-rays
are below the cooling frequency), the evolution of the pre-jet break
optical and X-ray afterglow,  and the sharp
turn-on observed for some afterglows.
Recent observations of short bursts, which are expected to be interacting
with a constant density medium, provide a check on the standard afterglow
model.
Although radio observations do not support the constant density model
for long bursts in some cases, the evidence for constant density interaction
is strong.
The most plausible way to produce such a medium around a massive star
is to shock the progenitor
wind.
This requires a smaller termination shock than would be expected, possibly
due to a high pressure surroundings, a high progenitor velocity,
 or the particular evolution leading to a GRB.
However, the need for the termination shock near the deceleration radius
cannot be plausibly accomodated and may indicate that some long bursts
have compact binary progenitors and explode directly into the interstellar
medium.

\end{abstract}


\end{frontmatter}


\section{Introduction}
\label{intro}

Observations of the long duration gamma-ray bursts 
(GRBs), which are the primary topic here,
suggest that they are associated with the deaths of massive stars.
One line of evidence for this has been their association with supernovae (SNe).
\cite{WB06} list 11 cases that are good candidates for SNe associated
with bursts.
However, 2 recent long bursts have shown no evidence for a
SN to faint levels \citep{Fyn06}.
Another line of evidence is the apparent association of the sites of long
bursts with regions of active star formation.
\cite{Fru06} found that the positions of GRBs on galaxies are more
concentrated to the brightest pixels than are core collapse supernovae,
suggesting that the GRBs are associated with more massive stars 
than are most core collapse supernovae.

These properties imply that the progenitors of most explosions
are Wolf-Rayet (WR) stars.   Expectations for the surroundings of WR 
stars are discussed in Section \ref{surr}.
Normal SNe Ib/c (without a GRB connection) are also expected to interact
with the surroundings of WR stars.
The properties of the interaction, as observed at radio and X-ray wavelengths,
can provide insight into the GRB case (Section \ref{csinter}).
The SNe Ib/c that are associated with nearby, low luminosity
GRBs distinguish themselves from
the normal SNe Ib/c and are discussed in Section \ref{lowl}.
Afterglow emission gives us a prime method of determining the surrounding
medium and has been used since the discovery of afterglows to infer the
density.
This issue is examined in the light of recent observations in Section \ref{ag}.
A frequent deduction from the analysis of afterglows is that the surroundings
have a constant density.
In this paper, GRB is used to refer to the long duration GRBs.
The short bursts are discussed in relation to the long bursts
in Section \ref{sgrb}.   
Possible ways of producing a constant density surroundings
are discussed in Section \ref{const}.
Absorption lines in optical spectra provide another possible window
on the immediate surrounding of GRBs and are treated in Section \ref{abs}.
The various issues related to the surroundings
of GRBs are summarized in Section \ref{concl}.

\section{The Surroundings of Wolf-Rayet Stars}
\label{surr}
In this section, I consider the surrounding medium created by the free
wind from a WR star, as it is likely to provide the immediate environment
for the GRB.
At some point the wind is expected to transition to a region that
results from interaction with the surroundings; this possibility will
be considered in Section \ref{const}.
Typical parameters for a WR star wind are 
a mass loss rate $\dot M=10^{-5}\ml$
and a wind velocity $v_w=10^3$ km s$^{-1}$.
The wind density $\rho_w=Ar^{-2}$, where $A=\dot M/4\pi v_w$, is the
critical parameter for a high velocity interaction and I characterize
it by $A_*=A/5\times 10^{11}{\rm~g~cm^{-1}}=(\dot M/10^{-5}{\ml})
(10^3 {\rm~km~ s^{-1}}/v_w)$.
For Galactic stars, \cite{NL00} listed mass loss parameters for 
64 WR stars, yielding an $A_*$ range of 0.07 to 7.4.
The lowest density winds are produced by WO stars, because of their high
values of $v_w$, up to $5500\kms$.
These results are based on 
\cite{NCW98}, who determined clumping-correcting radio mass loss rates,
noting that the effect of clumping is small at the stellar surface,
grows to a maximum at $\sim5-10 R_*$, and again becomes small in
the outer wind because of the expansion of clumps at the local
sound speed.

A  difference of the progenitors of GRBs with Galactic WR
stars is that the GRB progenitors probably have lower metallicity.
\cite{Mod07} found that the nearby GRB/SNe are in regions that are
systematically more metal poor than the regions containing core collapse 
supernovae.
Metallicities were determined from emission line regions close to the
explosions.
The metallicities could be a factor of $\sim 6$ smaller than solar.
For more distant GRBs, it is not clear whether the GRBs have lower metallicities
compared to other galaxies at a similar redshift $z$, but 
metallicities $Z\sim 0.05-0.5~Z_{\odot}$ are indicated.
In recent years, it has been found that heavy elements around the Fe peak
play a role in driving the winds from WR stars, so that their
mass loss rates are $Z$ dependent.
In this metallicity range, mass loss rates from WC stars vary
as $\sim Z^{-0.6}$ \citep{Cro06}, suggesting that values of $\dot M$ for GRB progenitor
stars are lower than the rates for Galactic WR stars by a factor
of $2-3$, and the values of $A_*$ are lower by a similar factor.

Another issue is the possible asymmetry of the stellar wind.
Polarization studies of Galactic WR stars have generally shown an
undetectable amount of polarization, although $\sim20$\% show
a  polarization $\gsim0.3$\% that can be interpreted as a density contrast of
a factor of a 2--3 \citep{HHH98}.
However, asymmetry could be a significant factor for the small percentage
of WR stars that become GRBs.
A plausible distinguishing feature of the GRB WR stars is rapid rotation,
so that the central core is rapidly rotating.
For radiation driven winds, the higher gravitational acceleration on the
polar axis can lead to a higher radiative flux and mass loss rate
on this axis, although a lower temperature and higher opacity on
the equator favors equatorial mass loss \citep[e.g.,][]{M02};
the density contrast from pole to equator can be a factor of a few.
\cite{MM07} have suggested that higher mass loss along the polar
axis is needed so that the WR progenitor does not lose too much
angular momentum through its wind.

Overall, the additional effects to consider for GRB progenitors compared
to Galactic WR stars do not have a substantial effect on the expected wind density 
when compared to the large possible range of densities.

\section{Circumstellar Interaction of Normal Type Ib/c Supernovae}
\label{csinter}

The interaction of normal SNe Ib/c with their surroundings provides an
interesting case of comparison for the GRB case because the driving
force is generally understood for supernovae and there is synchrotron
emission resulting from the interaction as in the the GRB case.
In a normal supernova, the supernova shock wave accelerates through the
steep density profile at the outer edge of the star; the acceleration
stops when radiation can stream freely from the star.
The radiation accelerates the outer gas and the radiation dominated
shock front disappears. 
The shock wave re-forms as a viscous shock in the surrounding stellar
wind.
There is a shocked region bounded by a reverse shock on the inside and
a forward shock on the outside.
The reverse shock is in the steep outer power law portion of the
supernova density profile.
The shock fronts are plausible sites of particle acceleration;
however, the material entering the reverse shock front has a very
low magnetic field because of the supernova expansion, so there
is some question of the efficiency of particle acceleration at 
that site.

The deceleration of the ejecta by the surrounding medium gives rise
to a Rayleigh-Taylor instability and a turbulent region in the shocked
layer.
The magnetic field can be built up in this region, although it is not
clear that the efficiency is high.
In numerical simulations, \cite{JN96} found that the field is strongest
on the smallest scales.
The energy density in the field was limited to $\sim0.3$\% of the turbulent energy
density, but the result was limited by the numerical resolution.
Another possible source of magnetic amplification is related to instabilities
in the collisionless shock waves \citep[e.g.,][]{Bel04}.

The basic hydrodynamic model of steep power law ejecta driving an interaction
shell into a surrounding stellar wind can reproduce the observed radio
emission from SNe Ib/c if some fraction of the postshock energy density
goes into relativistic particles and magnetic fields, and synchrotron
self-absorption (SSA) is important at early times \citep{Che98,CF06}.
In this situation, the peak flux of the radio emission gives information
on the radius, and thus the velocity, of the radio emitting region.
Figure 1 shows observed peak fluxes and ages of radio supernovae.
The dashed lines give the velocities of the radio emitting
regions if the peak is due to SSA.
If another absorption process, such as free-free absorption, is dominant,
the velocity inferred from the turn-on is lower than the actual velocity.
In Fig. 1, it can be seen that the SNe II have lower inferred velocities,
which is both because of the relatively low velocities in SNe II and the
importance of free-free absorption.
There is a large range in peak luminosity for SNe II, which can be primarily
attributed to a range in circumstellar density.
Multiwavelength observations confirm the large range in density.

\begin{figure}
  \includegraphics[height=.3\textheight]{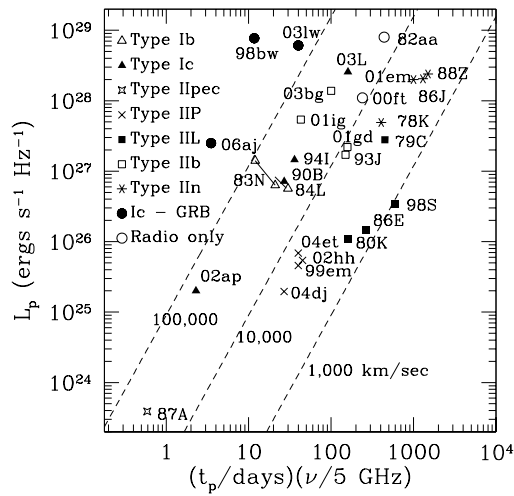}
  \caption{Peak radio luminosity and corresponding age for well-observed
core collapse supernovae.  The dashed lines give curves of constant expansion
velocity, assuming synchrotron self-absorption at early times
(updated version of Fig. 4 in \cite{Che98}).}
\end{figure}

The normal SNe Ib/c show systematically higher velocities than the SNe II
(Fig. 1), which can be attributed to some combination of  higher peak velocities
at the time of shock breakout, higher mean ejecta velocities because of lower
ejecta mass, and lower deceleration because of lower circumstellar density.
There is again a large range in peak luminosity; this may be due to a range
in circumstellar density, but in this case there is no independent evidence
for a range in density.
If the supernovae have similar efficiencies for the production of synchrotron
radiation and a range of circumstellar density that is comparable
to that around Galactic WR stars, the observed luminosity range can approximately
be reproduced if $\epsilon_B\approx \epsilon_e\approx 0.1$ \citep{CF06}.
The required magnetic field is high and cannot be produced by compression
of the wind magnetic field (this would require a wind energy flux that was
completely dominated by the magnetic field).

There are a number of X-ray observations of SNe Ib/c, but the data are much less
extensive than at radio wavelengths \citep[][and references therein]{CF06}.
The observed luminosities are higher than expected from thermal emission from
interaction with a normal WR star wind, so that a nonthermal mechanism is
indicated.
Near maximum light, inverse Compton scattering of photospheric photons with 
relativistic electrons is a possibility \citep{BF04,CF06}.
At later times, inverse Compton emission fades because of the low
supernova luminosity, and synchrotron
emission is the most plausible nonthermal mechanism.
However, an extrapolation of the radio synchrotron emission falls below
the observed X-ray emission, especially when synchrotron cooling of the
radiating electrons is taken into account.
\cite{CF06} suggested a model of particle acceleration in a cosmic ray dominated
shock front so that the particle spectrum flattens to high energy.
At low energy, the particle spectrum is relatively steep, with energy
index $p\approx 3$, in accord with radio observations of SNe Ib/c.
At high energies, the spectrum becomes flat.
This spectrum results in fairly flat evolution of the X-ray emission,
while the radio emission decreases.

\section{Low luminosity, nearby GRB-SNe}
\label{lowl}

Figure 1 shows that the 3 low luminosity GRB/SN events have higher velocities
of the radio emitting regions than the normal SNe Ib/c.
Their positions in the figure suggest semi-relativistic velocities.
The high luminosity nearby event GRB 030329 had a 5 GHz luminosity of
$5\times 10^{30}$ erg s$^{-1}$ Hz$^{-1}$ on day 10
\citep{Ber03}, indicating highly relativistic
motion in this case.
For the lower velocity cases, application of the synchrotron theory 
described by \cite{CF06} yields
mass loss densities $A_*(\epsilon_B/0.1)$ of 0.1 (SN 1998bw), 1.6 (SN 2003lw),
and 0.02 (SN 2006aj).
The theory is nonrelativistic, but should still yield approximate results
considering that these objects indicate only semi-relativistic motion.
The density inferred for SN 2006aj is low because of the early turn-on
\citep[see also][]{WMC07}.

An important issue for the 3 low luminosity events is whether some of the observed
phenomena can be explained by the supernova, or whether a central engine
is needed, as in the case of normal GRBs.
A supernova explanation means that emission associated with the interaction
of the fast, outer supernova ejecta can explain the observations.
Expectations for the supernova case depend on the ejecta mass and energy
for the explosions.
Optical observations of the 3 supernovae were extensive and there are
results on the supernova parameters: $10\Msun$ and $50\times 10^{51}$ ergs for SN 1998bw,
$13\Msun$ and $60\times 10^{51}$ ergs for SN 2003lw,
and $2\Msun$ and $2\times 10^{51}$ ergs for SN 2006aj \citep{Maz06a,Maz06b}.
The supernova properties of SN 2006aj were closer to the normal Ic
SN 2002ap than to the bright SN 1998bw, and its inferred mass and energy
would be incapable of producing the high velocity inferred from the radio
emission.
The implication is that the radio emission is related to a central engine.
The radio observations of SN 2002ap suggest a low wind density in this case (Fig. 1),
which appears to also apply to SN 2006aj.
For SN 1998bw and SN 2003lw, the large supernova energy allows the possibility
of a supernova origin for the radio emission.
\cite{TMM01} have discussed such a model for SN 1998bw.

SN 2006aj showed a thermal X-ray component over the first few 1000 sec
that has been interpreted by \cite{Cam06} and \cite{WMC07} as shock
breakout emission.
The temperature was constant at $\sim 0.17$ keV during this time.
The radiated energy in the thermal component was 
$\sim2\times 10^{49}$ ergs \citep{Cam06,Li07}. 
This is several orders of magnitude larger than would be expected from
shock breakout from a WR star, assuming no effect of the WR star wind
\citep{MM99}; in addition, the duration of the shock breakout emission 
in this case would
be determined by light travel time effects, yielding a timescale $\sim 10$ s,
much less than observed.
\cite{Cam06} addressed this issue by considering the progenitor star to
be surrounded by a dense WR star wind, with $A_*\approx 20$.
This wind density conflicts with that deduced from the radio emission, but
the radio emission is at later times and it is possible that there was a phase
of dense mass loss just before the explosion.
With the dense wind, the photosphere is formed at $r\approx 5\times 10^{12}$ cm.
The corresponding light travel time, 200 sec, is still less than the duration
of the thermal component, so \cite{Cam06} and \cite{WMC07} appeal to an asymmetric progenitor
structure to lengthen the timescale.
Another issue is the total energy emitted in the thermal component, considering
the fairly low energy explosion estimated for SN 2006aj mentioned above.
\cite{Li07} considered supernova shock breakout in a wind and found that the energetics
present a problem for SN 2006aj.
However, \cite{WMC07} attributed the emission to the breakout of a mildly
relativistic shell; it is possible that such shell ejection could be generated
by a central engine.
It thus appears that the early X-ray emission from SN 2006aj cannot be accounted
for by the supernova and activity of the central engine is needed.
Whether the thermal emission can be explained by breakout emission remains uncertain.
One issue is whether the constant temperature emission can be produced if the
progenitor is highly asymmetric.
More detailed modeling of the emission is needed.

\section{GRB Afterglows}
\label{ag}

As discussed in Section \ref{surr}, the immediate surroundings of a long GRB is 
expected to be the wind from the progenitor star and one would expect the
afterglow to reflect interaction with such a surroundings.
In the time before a jet break occurs, there are clear differences between
evolution in a wind medium and in a constant density (often referred to
as ISM for interstellar medium) \citep{CL00}.
The cooling frequency, where the synchrotron cooling time equals the age,
increases as $t^{1/2}$ in the wind case, but decreases as $t^{-1/2}$ in the
ISM case.
This frequency typically occurs between optical and X-ray wavelengths, giving
the expectation that the flux should drop more rapidly with time at optical
wavelenths than in X-rays for the wind case.
The opposite is true for the ISM case.
The peak flux, $F_{\nu m}$, at the typical frequency $\nu_m$, is lower at lower
frequencies $\propto \nu^{1/3}$ in the wind case, but is constant in the
ISM case.
This effect can be best observed at radio wavelengths because of the
large range of wavelengths that they provide.
Finally, the synchrotron self-absorption, $\nu_a$, drops as $t^{-3/5}$ in
the wind case, but is constant in the ISM case.
Again, radio observations are generally needed.

The application of these differences to observed light curves is complicated
by jet effects, which were generally found to occur at early times ($<3$ days)
for bursts found during the {\it BeppoSAX} era.
Models of the deceleration of jets have shown some features that are not
present in the simple models \citep{Gra07}, so there is uncertainty in the
interpretation.
Overall, detailed models of afterglows observed during the {\it BeppoSAX} era
generally prefer interaction with a constant density medium, 
e.g., \cite{PK02}
who found that wind interaction was preferred for just 1 burst (GRB 970508) out of 10.
However, \cite{Star07} recently were able to
constrain the circumburst medium for 5
{\it BeppoSAX} sources, finding that 4 were consistent with a wind and
1 (GRB 970508) was consistent with ISM.
One difference with the analysis of \cite{PK02} is that radio data
were not included.
In the {\it Swift} era, there have been excellent data on early X-ray
afterglows, but there have been few extensive multiwavelength data sets.
A reason for this is that the greater sensitivity of {\it Swift} compared to
the previous GRB satellites, so the bursts are fainter in multiwavelength
observations.
In particular, there have been few radio light curves, although the radio
emission can provide important constraints, as described above.

Another aspect of {\it Swift} bursts is that the jet break often appears
fairly late in the evolution, if at all.
To some extent, this can be attributed to the discovery of lower
luminosity bursts.
An advantage of such bursts is that there is the possibility of 
using the distinguishing properties of wind vs. ISM models discussed above.
A well-observed burst is GRB 050820A, which did not show a jet break
until an age $\gsim 17$ days \citep{C06}.
During the pre-jet break period, \cite{C06} found that the X-ray afterglow
declines more rapidly than the optical afterglow, which is an indicator
of ISM interaction.
However, an ISM model that is consistent with the optical and X-ray
properties overpredicts the radio emission.
\cite{C06} expect a radio flux of $5$ mJy on day 7, but observe
a flux of 0.1 mJy.
The low radio flux is consistent with a wind interaction model.
Thus the situation is ambiguous.

In addition to the finding of late X-ray breaks in {\it Swift} bursts,
optical observations sometimes show a break when none is present
at X-ray wavelengths \citep[e.g,][]{Mon06}.
This causes some uncertainty about the nature of the optical
break and may indicate that the X-ray and optical emission come
from different regions.

Although multiwavelength modeling provides the best constraints
on afterglow models, the large set of X-ray light curves and spectra observed
with {\it Swift} can be used for comparison with the expected ``closure
relations'' for ISM and wind models.
In a study of {\it Swift} bursts from the first 6 months of operation,
\cite{Zha06} found that all the bursts were consistent with ISM interaction.
However, when $\nu_c$ is below X-ray frequencies, the afterglow evolution does not
depend on the density profile, limiting the number of objects for which an
interesting result can be obtained.
In a more recent study of 30 sources, \cite{Pan07} found that  2/3 are consistent
with $\nu_c$ below X-ray frequencies so the afterglow evolution does not
depend on the density profile, 25\% are consistent with ISM evolution,
and 10\% are consistent with wind evolution.

The afterglow modeling described above applies to the blast wave phase of
evolution in which the ejecta have been decelerated by the surrounding
medium.
The development of rapid response optical/infrared telescopes has given the
possibility of making observations before the blast wave phase has been established.
The REM telescope may have observed GRB 060418 and GRB 060607A during
the onset of the afterglow phase \citep{Mol06}.
Before deceleration, the observed flux is expected to increase as
$t^3$ (ISM) or $t^{1/3}$ (wind) \citep{Mol06,JF07}.
The observed increases for the 2 bursts are consistent with a $t^3$
dependence and inconsistent with $t^{1/3}$, implying a constant density
interaction.
An estimate of the radius at which deceleration occurs is $10^{17}$ cm
\citep{Mol06}, showing that the constant density medium must extend in
to at least this radius.
Another case where a sharp turn-on of the afterglow may have been
detected is GRB 060206 \citep{Stan07}.
If this interpretation of the rise phases is correct, the further
light curves should be of the ISM type (or something more complex),
and not of the wind type.
The available information on these afterglows does not seem
to fit the simple models.
Another burst with early optical observations is GRB 050801, which
\cite{Ryk06} found to have a flat flux evolution from $20-250$ s.
The early flat evolution is roughly consistent with wind evolution,
but the later afterglow evolution is consistent with ISM interaction,
and not with wind interaction.
It is possible that the burst made a transition from wind to constant
density surrounding medium, but this would have to occur close to
the deceleration radius.
Although there are tantalizing clues from the early optical observations,
they cannot be clearly interpreted in terms of the standard models.

One of the main findings during the {\it Swift} era is that X-ray
afterglows are more complex than previously recognized, showing
a variety of flaring behavior and an early plateau phase.
Detailed observations of optical afterglows have also shown
complex evolution \cite[e.g.,][]{Dai07}.
These observations point to later energy addition to the GRB
blast wave than is assumed in the standard models.
Our lack of knowledge of the expected form of the energy addition
(unlike the supernova case) limits our ability to precisely
deduce the nature of the surrounding medium.
However, there is currently evidence for afterglow evolution
in both wind and constant density media.

\section{Short vs. Long GRBs}
\label{sgrb}

According to present thinking, the long GRBs are explosions in massive
stars, while the short bursts result from the mergers of compact objects.
These 2 progenitor types can be expected to have different environments:
the long bursts occurring in the mass loss of the progenitor stars
and the short bursts in the surrounding ISM.
A comparison of the afterglows for the two types of bursts can then
give an indication of whether the interpretation of long bursts interacting
with a constant density medium is correct and is not the result of an
effect such as the variation of microphysical parameters.

A good case is the analysis of GRB 051221A by \cite{Sod06}.
The time of an apparent jet break was 5 days, so there was significant
evolution in the pre-jet break regime.
The X-ray afterglow decline was characterized by $\alpha=-1.06\pm0.04$,
which, together with the X-ray spectral index, was consistent with
evolution in the cooling regime.
The flatter optical/X-ray spectral index and the flatter evolution at
optical wavelengths were roughly consistent with evolution in a constant density
medium with $\nu_c$ between optical and X-ray wavelengths, assuming
a standard afterglow model.
This result gives confidence in the application of the standard model
to those long bursts with the deduction that they are expanding into
a constant density medium.
The similarity between the short and long burst afterglow evolution
indicates that the apparent ISM interaction is not due to wind interaction
with a particular evolution of the microphysical parameters.

The observations of the short bursts are generally consistent with interaction
with a low density ISM.
In the case of GRB 051221A, \cite{Sod06} deduced a density $\sim10^{-3}$ cm$^{-3}$.
The lack of observable X-ray afterglows for some short bursts may be due
to a very low surrounding density, $\lsim10^{-5}$ cm$^{-3}$ \citep{Nak07}.

\section{Producing a Constant Density Surrounding Medium}
\label{const}

The evidence from afterglow modeling for constant density media around long
GRBs has stimulated interest in producing such a medium around a massive
star.
The most plausible way of doing so is the medium produced downstream of
the termination shock in the stellar wind \citep{Wij01}.
This region has a roughly constant pressure because the sound speed is
higher than the systematic velocities over most of the volume.
In addition, the velocities are sufficiently high to make the flow steady
over much of the volume, so that conservation of entropy with radius
leads to a constant density region.
The radius of the termination shock, $R_t$, can be estimated from the pressure
generated at the shock
\begin{eqnarray*}
R_t=5.7\times 10^{19}\left(v_w\over 10^3\kms\right)   \\
\left( p/k\over 10^4 {\rm~cm^{-3}~K}\right)^{-1/2} A_*^{1/2} {\rm~cm}
\end{eqnarray*}
where $p$ is the pressure in the shocked wind and $k$ is Boltzmann's constant.

A general problem is that the value of $R_t$ needed to explain the
afterglow observations is smaller than expected around a typical WR star.
Afterglow models require that the termination shock be at a radius
$\lsim 2\times 10^{17}$ cm in some cases \citep{CLF04}.         
One factor is the reduced value of $\dot M$ because of the low metallicity
of the progenitor \citep{Wij01}.
As discussed in section \ref{surr}, this reduces $\dot M$ by a factor $\sim3$, which
reduces $R_t$ by up to $\sim 2$.
Although this helps the problem, more is needed.
The other possibility is increasing the  pressure, which can be
accomplished by interaction with a dense ambient medium, high ram pressure
due to motion of the progenitor star, or high pressure of the ambient medium
\citep{van06,CLF04}.
The finding of \cite{Fru06} that GRBs occur in regions of strong star formation
may be consistent with the presence of a high interstellar pressure, but
this needs to be verified in more detail.

In the pre-burst models of \cite{van06}, the wind from the WR star sweeps out the 
dense red supergiant wind from a previous evolutionary phase.
However, there is increasing evidence for supernovae occurring soon after
the loss of the the H envelope in dense mass loss; an example is SN 2001em 
\citep{CC06}.
GRBs might preferentially occur in such an explosion because there is
less opportunity for loss of angular momentum in the WR wind.
The dense mass loss then provides a wall for shocking the WR wind.

A possible problem for the shocked wind explanation of the constant density medium
is the radial range over which it is required.
Observations of some GRBs requires that the  outer extent of
the shocked wind be
$\gsim 2R_t$, which rules out some wind interaction models \citep{CLF04}.
While some shocked wind models are consistent with these results, the lack of
evidence for interaction with the region inside or outside the shocked
WR wind is a possible problem.
In particular, one would expect an interaction with a freely expanding
wind followed by a transition to constant density medium.
Early work on this transition indicated that there would be increase in
emission when the shock was traversed \citep{Wij01,PW06},
but \cite{NG06} find that there is no bump in the GRB light curve at this point.
In any case, there should be a transition from the self-similar blast
wave evolution in a wind medium to evolution in a constant density medium.

The evidence for the turn-on of some afterglows in an ISM medium
(Section \ref{ag}) is a problem
for this model.  
The deceleration of the GRB ejecta in a wind occurs at a radius
\[
R_{dec}\approx 4.0\times 10^{15}E_{53}\Gamma^{-2}_{0,2}A_*^{-1}\quad {\rm cm},
\]
where $E_{53}$ is the isotropic blast wave energy in units of $10^{53}$ ergs
and $\Gamma^{-2}_{0,2}$ is the initial Lorentz factor of the GRB
ejecta in units of $10^2$ \citep{PK00}.
The pressure needed to have the termination shock occur at or within this
radius cannot plausibly be attained \citep{van06}, so a massive star progenitor
may not be viable in these cases.
A possibility is that some long bursts have compact binary progenitors
and interact directly with the ISM.
\cite{KOD07} have suggested the merger of neutron stars and white dwarfs as
possible long burst progenitors.
These events would generally be associated with active star formation (but not
always) and would not be accompanied by a supernova.

\section{Clues from Absorption Lines}
\label{abs}

Possible information on the immediate surroundings of GRBs comes from
absorption lines observed in the optical spectra.
The mass loss processes leading up to explosion have the possibility of
creating absorption features in the spectrum; in particular, the free wind
is expected to have a velocity as high as $5000\kms$, which can be
distinguished from typical velocities in the host galaxy.
A redshift $z \gsim 2$ is needed so that the strong ultraviolet resonance
line transitions are shifted into the optical.
A case of special interest has been GRB 021004 ($z=2.3$), which showed a
number of absorption line systems with velocities up to
$3000\kms$ in observations by \cite{Sch03} and \cite{Mir03}, who concluded that
the features were likely to be circumstellar.
Higher resolution observations have shown that the $-3000\kms$ line system
can be separated into $-2700\kms$ and $-2900\kms$ systems \citep{Fio05,Chen07}.
The problem with the circumstellar 
hypothesis is that the radiation field of the GRB
is expected to completely ionize the gas around the burst to a substantial
radius \citep{Laz06,Chen07}.
\cite{Mir03} suggested that the high velocity features are due to clumps initially
at some distance from the burst that are radiatively accelerated by the burst.
However, it is not clear whether coherent acceleration by the radiation field
could take place.
An alternative point of view is that the $3000\kms$ features are due to the
freely expanding WR star wind, which has the advantage of naturally explaining
the observed velocity.
However, there is still the problem of the strong ionization.
\cite{Laz06} deal with this problem by suggesting that the WR wind termination
shock is out at 100 pc, where the free wind is not completely ionized.
This value of $R_t$ is in the opposite direction to what is generally needed
for GRB afterglows (Section \ref{const}) and requires an unusually {\it low} surrounding pressure, in addition to the low surrounding density assumed
in the model of \cite{Laz06}.

Another problem with the WR star wind hypothesis is that the lines observed
in spectrum of GRB 021004 include H lines, which are generally not expected
in WR star wind at the end of the star's life.
The presence of H would not be a problem if the absorption is formed in
an intervening system.
\cite{Chen07} undertook a project to check whether absorption line systems in the
velocity range $1000-5000\kms$ were due to intervening systems or to the
progenitor winds.
The finding of one high velocity system out of 5 observed GRBs was consistent
with an intervening system.
They also argued that the high velocity systems observed in GRB 021004 could be
attributed to an intervening system, citing the presence of H I, C II, and Si II
together with the absence of excited C II or Si II as evidence.
Overall, it appears that absorption line observations are not likely to give
information on the medium that the GRB is exploding into.

One way to demonstrate a relation between the absorbing gas and the GRB would
be line variability due to the GRB radiation.
The C IV absorption features in GRB 021004 were measured over 6 days
but do not show any clear evidence for variability \citep{Laz06}.
\cite{Vre06} found evidence for  variability of Fe II and Ni II absorption
lines toward GRB 060418.
They showed that
ultraviolet pumping of Fe II and Ni II excited- and metastable-level populations by the GRB radiation is plausible, but their
model requires that the absorbing gas be $\gsim 1.7$ kpc from the GRB.
There is no significant Fe II or Ni II closer than this distance, presumably
because of ionization by the GRB radiation.

\section{Discussion and Conclusions}
\label{concl}

At the present time, there is a dichotomy between the supernovae and the GRBs.
In the supernova case, the driving force for the interaction with surroundings
is fairly well understood.
The shock acceleration through the outer parts of a star 
does not depend on the details of the central explosion.
Models for the nonrelativistic interaction with the surroundings of the
progenitor star are straightforward and give strong support for models of
interaction with a free wind.
For GRBs, the relativistic, collimated flow depends on the details of matter
and energy production by the central engine.
Observations of afterglows in the {\it Swift} era have shown that the ejecta
properties are probably crucial for understanding the early 
afterglow evolution.

The standard afterglow model is probably a better approximation at later times.
From the beginning of afterglow observations, the standard interpretation
of long burst afterglows have indicated that, in the majority of cases,
the evolution is consistent with interaction with a constant density medium
(except for \cite{Star07}).
This trend has continued in the {\it Swift} era.
The indications of a constant density medium include the evolution of the
X-ray afterglow when the cooling frequency appears to be above X-ray wavelengths,
and the evolution of the optical afterglow and its relation to the X-ray
evolution.
Radio observations do not support constant density interaction, and that is 
one of the main points against this picture.
More recent support for constant density interaction in some cases comes from
the sharp turn-on of optical afterglow emission.
Also, the consistency of the standard constant density afterglow model with
observations of short bursts gives confidence in the similar model when applied
to long bursts; short bursts are expected to be interacting with a constant
density ISM.
In other cases, there is evidence for interaction with a wind medium.

There is thus sufficient evidence for constant density interaction that
ways of producing such a medium must be considered.
The inner boundary of the constant density medium must extend in to
$\lsim2\times10^{17}$ cm.
This is larger than the radial distance usually sampled by supernova observations,
so there is not a clear discrepency between these cases.
However, a constant density around a massive star is most plausibly produced
by having the stellar wind pass through a termination shock, and the
required radius is smaller than would typically be expected for a massive
star wind at the end of its life.
There is thus interest in how to produce a small termination shock radius
for GRB progenitors.

One effect is the lower mass loss rate expected for the low metallicity
progenitors of the GRBs, but additional effects are needed to bring in
the radius.
One possibility is that the GRBs occur in regions of the ISM with a high
pressure, which reduces the value of $R_t$.  
Some support for this is that the bursts are observed to be more concentrated
to star forming regions than are supernovae \citep{Fru06}.
The action of stellar winds and supernovae from massive stars can produce
a high pressure.
Other possibilities are a high space velocity in a dense medium
or that the progenitor stars undergo a particular
evolution that results in a small $R_t$.
However, the observational evidence for interaction with a constant density
medium at the GRB deceleration radius is difficult to reconcile with a
shocked wind model and may indicate a compact binary progenitor in the ISM
for these cases.
In this situation, the GRB should not  be accompanied by a supernova.
The surroundings of GRBs may be indicative of various progenitor types.

\begin{ack}
I am grateful to Z.-Y. Li, C. Fransson and A. Soderberg for collaboration
on these topics, and to NASA grant NNG06GJ33G for support.
\end{ack}

\end{document}